\documentclass[dvipdfm]{aa}
\usepackage[dvips]{graphicx}
\usepackage{txfonts}
\usepackage{natbib}
\bibpunct{(}{)}{;}{a}{}{,}

\begin{document}

\title{Polarisation properties of Milky-Way-like galaxies}

\author{X. H. Sun\inst{1, 2, 3}
        \and W. Reich\inst{1} 
       }

\institute{Max-Planck-Institut f\"{u}r Radioastronomie, 
           Auf dem H\"ugel 69, 53121 Bonn, Germany\\
           \email{wreich@mpifr-bonn.mpg.de}
           \and
           Sydney Institute for Astronomy, School of Physics, The University of 
           Sydney, NSW 2006, Australia\\
           \email{xiaohui.sun@sydney.edu.au}
           \and
           National Astronomical Observatories, CAS, Jia-20 Datun Road, 
           Chaoyang District, Beijing 100012, China}
\date{Received / Accepted}

\abstract
{} 
{We study the polarisation properties, magnetic field strength, and synchrotron 
emission scale-height of Milky-Way-like galaxies in comparison with other 
spiral galaxies.
}
{We use our 3D-emission model of the Milky Way Galaxy for viewing
the Milky Way from outside 
at various inclinations as spiral galaxies are observed. 
We analyse these Milky Way maps 
with techniques used to obtain the strength of 
magnetic fields, rotation measures (RMs), and scale-heights of synchrotron 
emission from observations of resolved galaxies and compare the results
with the Milky Way model parameter. We also simulate a 
large sample of unresolved Milky-Way-like galaxies to study their 
statistical polarisation properties.
}
{When seen edge-on the synchrotron emission from the Milky Way has an 
exponential scale-height of about 0.74~kpc, which is much smaller than the 
values obtained from previous models. We find that current analysis methods
overestimate the scale-height of synchrotron emission of galaxies 
by about 10\% at an inclination of $80\degr$ and about 40\% at an inclination of 
$70\degr$ because of contamination from the disk.
The observed RMs for face-on galaxies derived from high-frequency polarisation 
measurements approximate to the Faraday depths (FDs) when scaled by a factor of 
two. For edge-on galaxies, the observed RMs are indicative of the orientation of 
the large-scale magnetic field, but are not well related with the FDs. Assuming 
energy equipartition between the magnetic field and particles for the 
Milky Way results in an average 
magnetic-field strength, which is about two times larger than the intrinsic value for 
a $K$ factor of 100. The number distribution of the integrated polarisation 
percentages of a large sample of unresolved Milky-Way-like galaxies peaks at 
about 4.2\% at 4.8~GHz and at about 0.8\% at 1.4~GHz. Integrated polarisation angles 
rotated by $90\degr$ align very well 
with the position angles of the major axes, implying that unresolved 
galaxies do not have intrinsic RMs.
} 
{
Simulated maps of the Milky Way Galaxy when viewed from outside at various 
inclination angles are the basis for a comparison with spiral galaxy maps. 
They are also helpful to check the accuracy of analysing methods of spiral galaxy observations.  
Unresolved Milky-Way-like galaxies are ideal background sources to investigate 
intergalactic magnetic fields. However, at 1.4~GHz they are typically polarised 
below about 1\% and hence a high instrumental purity of future facilities such 
as the SKA is required to observed them.
}


\maketitle

\section{Introduction}

Understanding the role of the magnetic field in spiral galaxies requires to
measure its properties and relation to other components of the interstellar 
medium. The small-scale magnetic fields of the Milky Way Galaxy can be studied 
in detail. However, properties of the large-scale field are difficult to 
recover, since the Solar System is located within the disk. 

Rotation measures (RMs) of polarised background sources are important probes of 
the large-scale regular magnetic fields in the Milky Way \citep[e.g.][]{btj03, 
bhg+07, tss09, vbs+11}. To obtain a very dense grid of 
RMs is the major aim of the Magnetism Key Science Project 
\citep{bg04} to be carried out with the future Square Kilometre Array (SKA) and 
the various pathfinder telescopes such as ASKAP \citep{jbb+07} at around 
1.4~GHz. At this frequency, however, intrinsic polarisation properties of the 
various classes of background sources are poorly understood. Recently, 
\citet{skbt09} investigated 23 resolved nearby spiral galaxies observed
 at 4.8~GHz and made a geometric model of the magnetic 
field to predict the statistical polarisation properties of unresolved galaxies 
at 1.4~GHz. Simulations with more realistic models are needed to achieve a 
better insight into the problem.

Polarisation observations have also been widely used to study magnetic fields of
 resolved nearby galaxies (see \citealt{bec05} for a review). RMs are usually 
obtained from polarisation measurements at short wavelengths, such as 
$\lambda$3.6~cm and $\lambda$6~cm, and are believed to indicate the intrinsic 
galaxy properties. Examples are the low-inclination galaxy NGC~6946 
\citep{bec07} and the nearly edge-on galaxy NGC~253 \citep{hkbd09}. 
Here it has been presumed that galaxies are optically thin for 
polarized emission at these wavelengths. However, 
we have recently found from the Sino-German $\lambda$6~cm polarisation survey 
of the Galactic plane \citep{srh+11}, that the polarisation horizon is about 
4~kpc towards the very inner Galaxy at $\lambda$6~cm, just a small portion of 
the Galaxy. This motivates us to investigate whether RM maps of nearby galaxies 
correctly show the intrinsic RMs of galaxies.  

The Milky Way Galaxy is a typical spiral galaxy, whose magnetic fields have 
been extensively studied by RMs of pulsars \citep[e.g.][]{hml+06,njkk08} and 
extragalactic sources as mentioned above, as well as by total 
intensity and polarisation all-sky surveys \citep[e.g.][]{wlrw06,trr08,phk+07}. 
By taking into account all relevant radio observations available, 
\citet{srwe08} and \citet{sr10} proposed 3D-emission models of the Galaxy. 
Although many parameters need to be refined, the models are the most robust 
today as they agree with of the observations of RMs of 
extragalactic sources, the all-sky total intensity map at 408~MHz \citep{hssw82},
the polarisation map at 22.8~GHz \citep{gow+11}, and the WMAP thermal emission
template \citep{gbh+09}. 

Based on the 3D-emission models, we mimic nearby Milky-Way-like galaxies 
by moving the Galaxy 
away to any distances and rotate it by arbitrary values. The large sample of 
simulated Milky-Way-like galaxies allows us to study the statistical 
properties of their 
integrated polarisation. We also checked whether the standard assumption of 
energy equipartition is relevant to estimate the strength of the average 
magnetic field, and to what extent the inclination of a galaxy affects the 
determination of its scale-height. 

The paper is organised as follows: the simulations of the Milky Way Galaxy and 
its edge-on view are described in Sect.~2, RMs, scale-height, and energy 
equipartition for resolved galaxies are investigated in Sect.~3--5. The 
integrated polarisation properties of a large sample of unresolved 
Milky-Way-like galaxies are studied in Sect.~6. We summarise our results in 
Sect.~7.

\section{Simulations}

\subsection{Galactic 3D-emission models}

Details of the Galactic 3D-emission models used in this paper were presented by 
\citet{srwe08} and \citet{sr10}. These models are the basis for the current 
investigations. They are able to reproduce the large-scale sky distribution of 
RMs of extragalactic sources as well as total intensity and polarisation 
all-sky surveys available at several frequencies. 

The magnetic field is separated into a large-scale regular field and an 
isotropic random field. The regular field consists of a disk and a halo 
component. The halo field is toroidal with opposite signs below and above the 
Galactic plane. The configuration of the disk field can be either axi-symmetric 
plus reversals following either concentric rings (ASS+RING) or the spiral arms 
(ASS+ARM), or bi-symmetric. Recent observations of numerous RMs 
in the Galactic plane by \citet{vbs+11} favor the ASS+RING field configuration,
 which will be used for the present simulations. The local regular magnetic 
field strength is 2~$\mu$G and its scale-height is 1~kpc. The strength of the 
random field is 3~$\mu$G elsewhere.      

The 3D-model of the thermal electron density we used is NE2001 \citep{cl02}. 
However, it is essential to adopt the recently revised scale-height of the 
diffuse ionised gas by \citet{gmcm08}, otherwise unrealistic parameters result 
as discussed by \citet{srwe08}. We therefore increased the scale-height of the 
thick disk component in NE2001 from 1~kpc to 1.83~kpc and changed the midplane
thermal electron density from $3.4\times10^{-2}$~cm$^{-3}$ to $1.4\times10^{-2}$~cm$^{-3}$.
More details were discussed 
by \citet{sr10}. For viewing the Milky Way Galaxy from outside, we removed all 
individual features such as clumps and voids included in the NE2001 model. The 
scale-height of the cosmic-ray electron density was set to 0.8~kpc, 
and the local density was taken as $6.4\times10^{-5}$~cm$^{-3}$. We briefly 
summarise the main model parameters used for simulations in the present paper 
in Table~\ref{modpar} and refer for details like spatial variations to 
\citet{srwe08} and \citet{sr10}.

\begin{table}[!htbp]
\caption{Model parameters.}\label{modpar}
\centering
\begin{tabular}{l|l}
\hline\hline
regular disk magnetic field & ASS+RING \\
                            & $z_0=1$~kpc, $R_0=10$~kpc\\
                            & local field strength: 2$\mu$G\\\hline
regular halo magnetic field & toroidal, asymmetry with plane \\ 
                            & maximum strength: 2$\mu$G\\\hline
random magnetic fields      & homogeneous, strength: 3$\mu$G \\\hline
cosmic-ray electron density & $z_0=0.8$~kpc, $R_0=8$~kpc, $\gamma=-3$\\\hline
thermal electron density    & NE2001 modified \\
                            & $z_0=1.83$~kpc for thick disk\\\hline
volume of the galaxy        & 40~kpc$\times$40~kpc$\times$10~kpc\\
\hline\hline 
\end{tabular}
\tablefoot{$z_0$: scale-height, $R_0$: radial scale-length, $\gamma$: energy 
spectral index which can be related to the spectral index of synchrotron 
emission as $\alpha=(\gamma+1)/2=-1$ ($I_\nu\propto\nu^\alpha$, 
with $I_\nu$ being the intensity).}
\end{table}

In our Milky Way model, we took the observed local synchrotron emission 
enhancement into account, which in turn reduced 
the large extent of the thick disk of previous models 
\citep[e.g.][]{pko+81,bkb85} to account for the observed emission at high latitudes. 
These earlier models were solely based on the all-sky 408~MHz map and the local
synchrotron enhancement was not established at that time. The problem is that
available low-frequency absorption data from optically thick \ion{H}{II} regions
\citep[e.g.][]{nhr+06,rcls99}, summarised by \citet{ft95}, did not constrain the 
3D-structure of the local enhancement very well. 
However, meanwhile there are more arguments against the existence of an extended intensive 
thick disk of the Milky Way.
The low percentage polarisation of the WMAP K-band (22.8~GHz) map \citep{gow+11}
requires a dominating random magnetic field. However, the regular magnetic field 
is constrained to a few $\mu$G by the RMs of extragalactic sources at high 
latitudes.
This in turn means a significant reduction of the cosmic-ray electron density to
account for the observed high-latitude synchrotron emission, which requires to suppress 
cosmic-ray electron diffusion out of the plane.  


The local emission enhancement was modelled as a synchrotron spheroid 
with uniform emissivity of 1~kpc radius, centered at $l=45\degr$, $b=0\degr$ 
with 560~pc distance from the Solar System \citep{sr10}. It is realised in the 
models by an enhanced cosmic-ray electron density, though one can equally 
achieve the same result by increasing the local random magnetic fields 
\citep{srwe08}. We kept the local enhancement, which has no influence on the 
global intensity distribution of the galaxies, as a mark of the position of 
the Solar System.

\subsection{Realisation of galaxies from the Milky Way Galaxy}

The simulations were performed with the \textsc{HAMMURABI} code 
\citep{wjr+09}. We revised the code to enable simulations of galaxies in a 
Cartesian coordinate system besides the standard HEALPix system \citep{ghb+05}. 

\begin{figure}[!htbp]
\centering
\resizebox{0.47\textwidth}{!}{\includegraphics[]{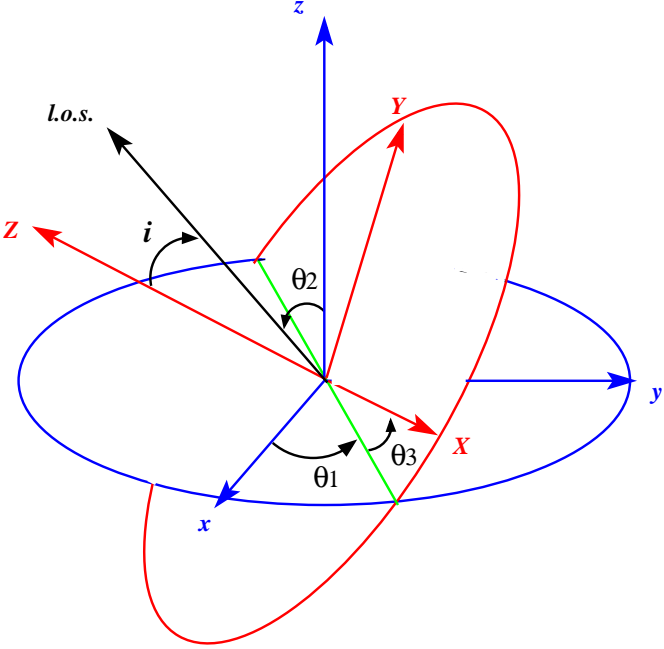}}
\caption{Rotation of the Galaxy from $xyz$-coordinates to $XYZ$-coordinates.}
\label{rot}
\end{figure}

To mimic a nearby spiral galaxy, the Galaxy was put at $l=l_0$, and $b=b_0$ at 
a distance $D$, and then rotated by Euler angles 
$(\theta_1,\,\theta_2,\,\theta_3)$. 
The process is explained in Fig.~\ref{rot}, where a galaxy at {\bf $xyz$} is 
rotated to $XYZ$. The origin of both coordinate systems corresponds to the 
centre of the galaxy, and $xy$ or $XY$ indicate the mid-plane of the disk of 
the galaxy. The Sun is located at $x=-8.5$~kpc, and $y=z=0$. The rotation is 
accomplished in three steps, 
\begin{itemize}
  \item around $z$ by $\theta_1$; 
  \item around new $x$ (green line in Fig.~\ref{rot}) by $\theta_2$; 
  \item around $Z$ by $\theta_3$. 
\end{itemize}
The inclination ($i$) of the new Milky-Way-like galaxy is determined by 
$\theta_1$ and $\theta_2$ as, 
\begin{equation}
\displaystyle{
\begin{array}{rcl}
\cos i&=&-\cos b_0\cos l_0\sin\theta_1\sin\theta_2+
         \cos b_0\sin l_0\cos\theta_1\sin\theta_2\\
      &&   -\sin b_0\cos\theta_2,
\end{array}}      
\end{equation}
where $0\leq\theta_1\leq2\pi$, and $0\leq\theta_2\leq\pi$. In the simulations 
presented in this paper, we set $l_0=90\degr$, $b_0=0\degr$. The inclination 
angle of the galaxy is then simplified as $\cos i=\cos\theta_1\cos\theta_2$. 

\section{Scale-height of the synchrotron emission}

\subsection{The scale-height of the Milky Way}

\begin{figure}[!htbp]
\centering
\resizebox{0.48\textwidth}{!}{\includegraphics[angle=0]{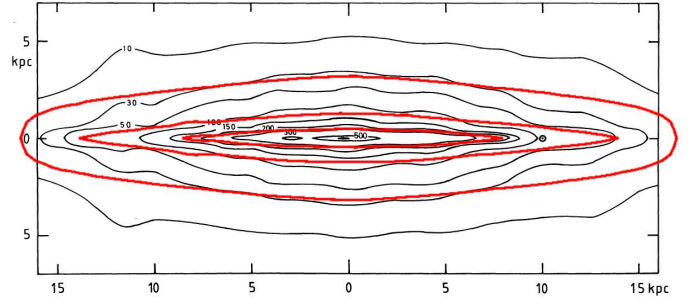}}
\caption{Edge-on view of the Milky Way Galaxy at 408~MHz. The black contours 
are from the model by \citet{bkb85}, and red contours with levels of 10~K, 
100~K and 300~K from outside to inside are from the present 3D-model.}
\label{i408}
\end{figure}

We simulated the synchrotron emission of the Milky Way Galaxy when seen edge-on 
at a distance of 5~Mpc. The sampling was chosen to be $0\farcm3$ to ensure 
that averaging effects are small. The total intensity of the Milky Way Galaxy 
seen edge-on at 408~MHz is significantly different from that modelled  
by \citet[][their Fig.~8]{pko+81} and \citet[][their Fig.~9]{bkb85} on the 
basis of only the 408~MHz all-sky survey \citep{hssw82}. The high-latitude 
emission was attributed to a thick disk component. \citet{pko+81} obtained a 
huge halo extending in $z$ up to about 10~kpc, and \citet{bkb85} obtained a 
slightly smaller value. We show an edge-on view of the \citet{bkb85} model in 
comparison with ours in Fig.~\ref{i408}, where the present $z$-extent is about 
a factor of two smaller. This difference results from the fact that the current 
simulations took into account the observed local synchrotron excess as reviewed 
in Sect.~2.1. 

\subsection{The scale-height obtained for inclined galaxies} 

\citet{dkwk95,dkw00} proposed a scheme to derive the scale-height of the 
synchrotron emission for highly inclined galaxies. The intensity profile along 
the major axis is projected to the $z$ axis by scaling the width by a factor 
$\cos i$. 
For an edge-on galaxy, the projection results in a delta-function at the centre.
The projected profile is then smoothed by using a Gaussian beam corresponding 
to the angular resolution of the observation. The smoothed profile is fitted by 
a Gaussian to yield a new half power beam width (HPBW), called the 
effective beam size. The observed disk profile of a galaxy is 
commonly assumed to be 
exponential ($\propto\exp(-|z|/z_{0,\,\rm e})$) or Gaussian 
($\propto\exp(-z^2/z_{0,\,\rm g}^2$)), which is convolved to the effective 
beam size. By fitting the newly convolved profile to the observation, the 
scale-heights ($z_{0,\,\rm e}$ or $z_{0,\,\rm g}$) can be obtained.
In this way, both the smoothing effects from the observations and the influence 
of the inclined disk can be accounted for. This procedure was subsequently used 
by many authors. Although this method is based on reasonable assumptions, there 
is no independent verification. 

\begin{figure}[!htbp]
\centering
\resizebox{0.48\textwidth}{!}{\includegraphics[angle=-90]{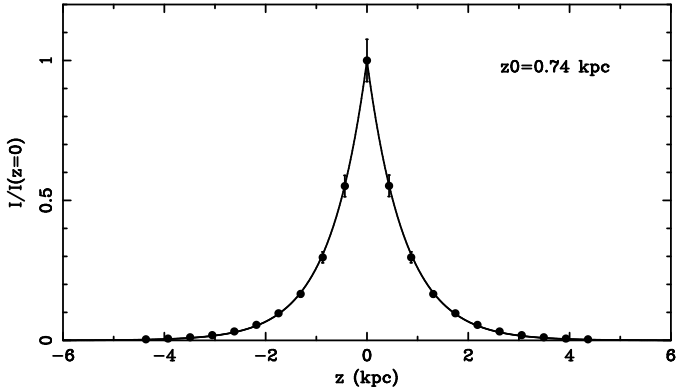}}
\caption{Normalised intensity distribution along $z$ for the Milky Way Galaxy 
when seen edge-on. The fitted exponential scale-height is $z_0$ = 0.74~kpc.}
\label{scalehc}
\end{figure}

We simulated the 4.8~GHz synchrotron emission from the Galaxy viewed edge-on 
at a distance of 5~Mpc with a sampling of $0\farcm3$, made stripe 
integrations perpendicular to the disk, 
and normalised the profile with the intensity maximum at $z=0$. The result is 
shown in Fig.~\ref{scalehc}. The distribution is well fitted by an exponential 
with a scale-height of 0.74~kpc, which is close to the value of 0.8~kpc for the 
cosmic-ray electron density in our model (see Table~\ref{modpar}). The 
intrinsic value is thus $z=0.74$~kpc, which one expects to obtain from 
observations at any inclinations and angular resolutions. 

\begin{figure*}[!htbp]
\centering
\resizebox{0.96\textwidth}{!}{\includegraphics[angle=-90]{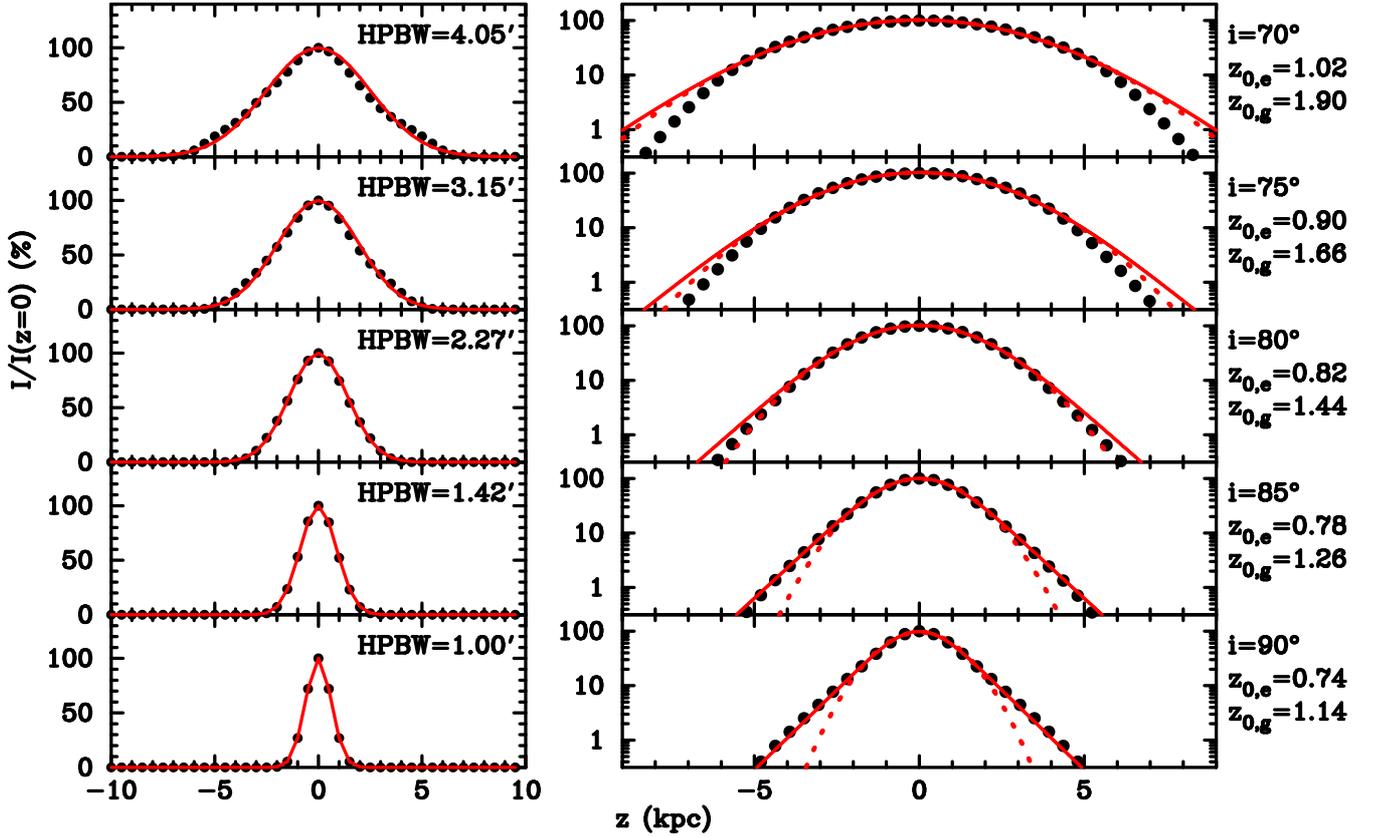}}
\caption{Average intensity distribution (shown by black dots) along $z$ for 
galaxies observed with a $1\arcmin$ beam. The projected and smoothed profiles 
used to obtain the effective HPBW as indicated is shown in the left column. 
Fits from exponential and Gaussian disks were shown by solid and dotted lines, 
respectively.}
\label{scaleh}
\end{figure*}

We simulated galaxies with inclinations from $i=90\degr$ to $i=70\degr$ 
in steps of $5\degr$, and smoothed the total intensity maps to an angular 
resolution of $1\arcmin$, which is about that of the Effelsberg 100-m 
telescope at 10~GHz. Stripe integrations were made in the same way as for the 
edge-on case shown in Fig.~\ref{scalehc} and the results are shown in 
Fig.~\ref{scaleh}.

For $i=90\degr$ the effective HPBW is $1\arcmin$ as expected. For smaller 
inclination angles the HPBW gradually broadens, which accounts for the 
contribution from the disk. For $i\gtrsim80\degr$, only models with 
exponential disks yield good fits (Fig.~\ref{scaleh}). This is expected because the 
input disk of the model is exponential. For small inclinations, the fits with an 
exponential or a Gaussian disk are both fairly good for intensities larger than 
about 10\% of the central value (Fig.~\ref{scaleh}). However, only models 
with exponential disks reproduce the intrinsic scale-heights reasonably.
 The scale-height is 
overestimated by about 10\% at $i=80\degr$ and about 40\% at $i=70\degr$. The 
Gaussian disks always produce too large scale-heights. For NGC 253 with 
$i=78\fdg5$, \citet{hkbd09} found $z = 1.7\pm0.1$~kpc. For this inclination,  
we got a scale-height 15\% larger than expected, which exceeds the quoted error,
 and indicates the inclination limit for the method. For galaxies with 
inclination angles below $i=70\degr$, the contribution from the disk masks the 
much fainter halo emission, so that a scale-height determination is not 
meaningful anymore.

For many galaxies, models with two exponential disks seem to give a better fit. 
Their typical scale-heights from $\lambda$6~cm observations are at the order of 
300~pc and 1.8~kpc \citep[e.g.][]{dk98,skdu11}. This is not the case for our Galaxy 
model, where individual source complexes in the disk are not included. We note 
that in most cases the difference between one- and two-disk models are based on 
intensities below about 10\% of the central values 
\citep[e.g. Fig.~6 in][]{dkw00}. More galaxies observed with a high 
signal-to-noise ratios are required for a thorough statistical study.

\section {Energy equipartition}

The assumption of energy equipartition or minimum energy assumption has been 
widely used to estimate magnetic fields in galaxies or supernova remnants.
\citet{bec05} gave a thorough review on this topic and a detailed discussion on 
its application. We followed \citet{bk05} to calculate the average magnetic 
field with the observed synchrotron total intensity and spectral index 
($\alpha$), and the so-called $K$ factor, which is the 
ratio between the proton and electron number density per unit energy. For 
spiral galaxies, the $K$ factor was traditionally set to 100 
\cite[e.g.][]{nb97} and more recent evidence favors this value 
\citep[see e.g.][]{hbkd11}.

\begin{figure}[!htbp]
\centering
\resizebox{0.45\textwidth}{!}{\includegraphics[angle=-90]{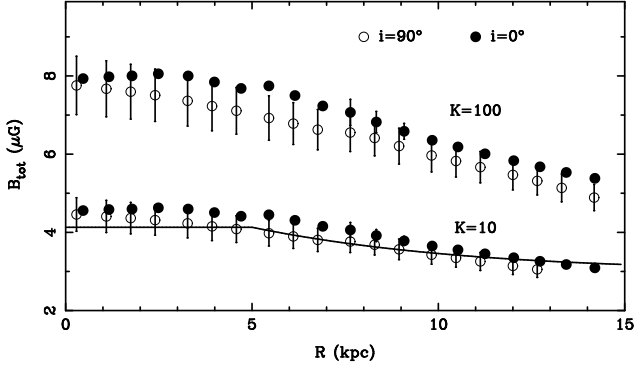}}
\caption{Average total magnetic field estimates based on energy equipartition 
for an edge-on (open circle) and a face-on (filled circle) view of the Galaxy.
The $K$ factor is  indicated. The total magnetic field strength of the model is 
shown by a solid line.} \label{beq}
\end{figure}

The true average total magnetic field in the disk, which we used in the model, 
can be represented as $B_{\rm tot}=\sqrt{B_{\rm reg}^2+B_{\rm ran}^2}$, 
where the regular field $B_{\rm reg}$ has an exponential shape at radii 
larger than 5~kpc and a constant value inside, and the random field 
$B_{\rm ran}$ is homogeneous with a variance of 3~$\mu$G \citep{srwe08}. 
The radial profile of the total magnetic field calculated this way is shown in 
Fig.~\ref{beq}.

We simulated the synchrotron emission for an edge-on and a face-on view of the 
Galaxy. In both cases we assessed the average total magnetic field based on 
equipartition \citep{bk05} with a path-length of 20~kpc and 2~kpc, respectively.
 For $K$ factor, we used the value 100, which is the currently best estimate. 
The results are shown in Fig.~\ref{beq}. The magnetic field strengths for both 
views are very similar, because the emission from the disk dominates the total 
intensity compared to the emission from the halo. At a radius of 
8.5~kpc, the field strength is consistent with the value 6$\pm$2~$\mu$G 
calculated by E. Berkhuijsen \citep[see][]{bec01} based on the emission 
in the plane from the \cite{bkb85} model (see their Fig.~6a).

The magnetic field strength estimated from energy equipartition with a $K$ 
factor of 100, however, overestimates the true field strength by a factor of 
nearly 2. To roughly reproduce the intrinsic total magnetic field, the $K$ 
factor should be about 10 (Fig.~\ref{beq}). However, as already 
pointed out above, $K=100$ seems to be well settled \citep{hbkd11}. The 
overestimated magnetic field strength indicates that the cosmic-ray particle 
energy is about four times 
larger than the magnetic energy in our model. The energy density of the magnetic 
field including both regular and random components is 
$\epsilon_B=(B_{\rm reg}^2+B_{\rm ran}^2)/8\pi\approx0.5\times10^{-12}$~ergs cm$^{-3}$ at $R=8.5$~kpc, and correspondingly the energy density of cosmic-ray 
particles is about $\epsilon_{CRE}=2.0\times10^{-12}$~ergs cm$^{-3}$. 
The local total pressure from cosmic rays and magnetic field is 
$P_{CRE+B}=\epsilon_B+\frac{1}{3}\epsilon_{CRE}\approx1.2\times10^{-12}$~dyn cm$^{-2}$. As summarised by \citet{cox05}, to support the gas against the gravity, 
the local non-thermal pressure from cosmic rays, magnetic fields and gas 
motions all together should be about $2.8\times10^{-12}$~dyn cm$^{-2}$. The 
dynamic pressure is therefore about $1.6\times10^{-12}$~dyn cm$^{-2}$, which 
dominates the pressure from magnetic fields. Recent MHD simulations by 
\citet{hjm+12} show that the magnetic field does not play an important 
role in supporting the gas. 

\begin{figure*}[!htbp]
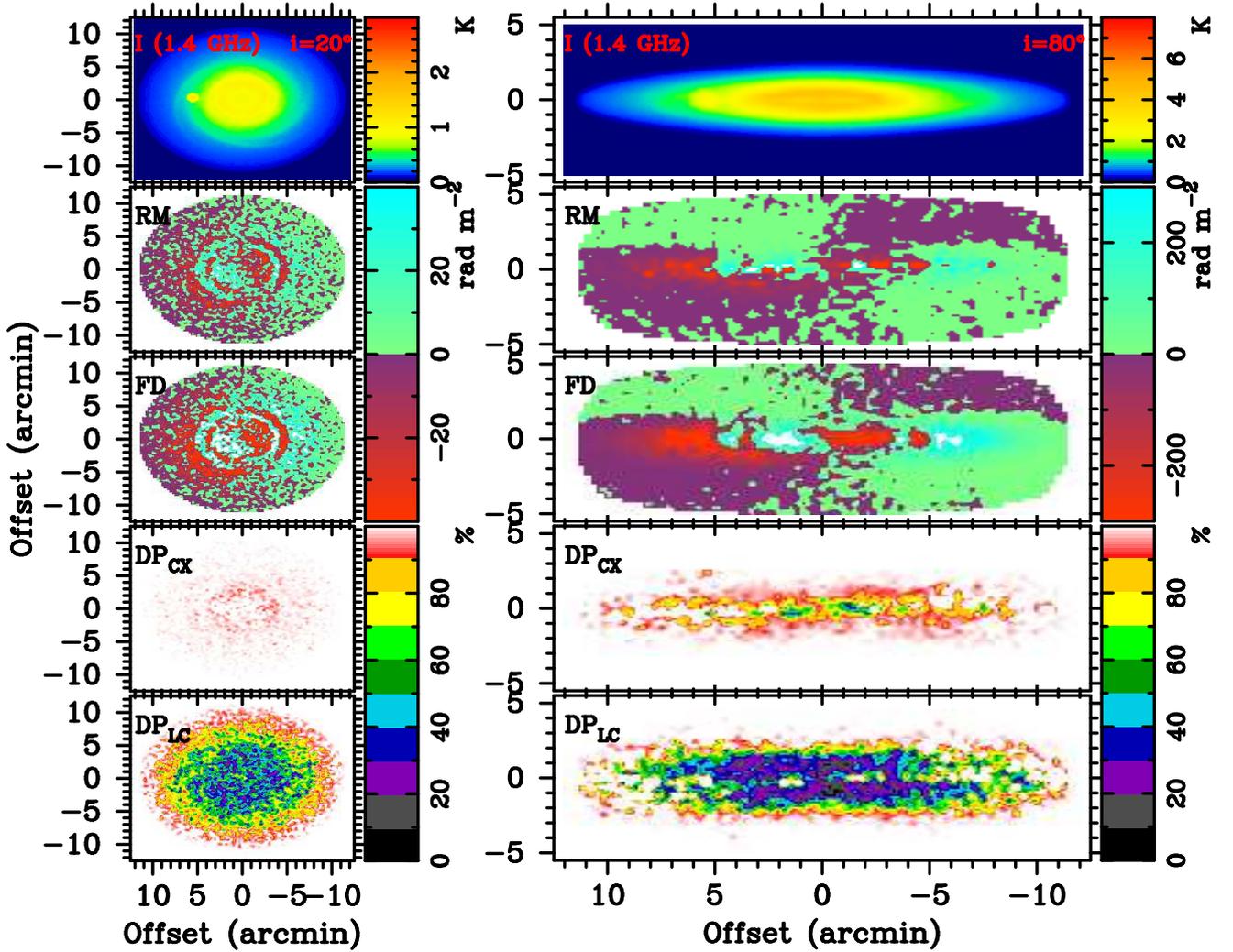

\centering
\resizebox{0.368\textwidth}{!}{\includegraphics[angle=-90]{dp.rm.nB.70.ps}}
\resizebox{0.57\textwidth}{!}{\includegraphics[angle=-90]{dp.rm.nB.10.ps}}
\caption{Total intensity (I) at 1.4~GHz, rotation measure (RM) calculated 
from 4.8~GHz and 8.4~GHz simulations, Faraday depth (FD), and depolarisation 
maps between 4.8~GHz and 8.4~GHz (DP$_{\rm CX}$) and between 1.4~GHz and 
4.8~GHz (DP$_{\rm LC}$) for two Milky Way galaxies with inclination angles $i=20\degr$ 
and $i=80\degr$, respectively.}
\label{rmmap}
\end{figure*}

It is clear that the equipartition magnetic field strength does not agree with 
the true value obtained directly from the model. This is not unexpected, as 
we know that our input model does not satisfy energy equipartition. In the same 
sense, it is also very uncertain whether spiral galaxies generally conform to 
the energy equipartition. It is thus questionable whether magnetic field 
estimates based on energy equipartition are reliable.

One may ask if it is possible to use the equipartition value for the magnetic 
field and adapt the cosmic-ray density accordingly to model the Galaxy. We have 
tested that and could not reach an agreement with the observations. A higher 
regular magnetic field will increase the RMs of pulsars and extragalactic 
sources, which could only be compensated by decreasing the thermal electron 
density. The latter, however, has been constrained by pulsar dispersion 
measures. Increasing the random field component will increase the total 
synchrotron emission and hence the cosmic-ray electron density must be reduced. 
This consequently reduces the polarisation intensity. To reproduce the 
observations, the regular magnetic field has to be increased, which is 
difficult as explained at the beginning. 

In case the Milky Way is not exceptional, the equipartition magnetic field strength 
estimated for $K=100$ might also be lower for other galaxies.
The reduction by a factor of two we found for the Milky Way corresponds to a magnetic energy 
density decrease by a factor of four and has a large influence 
on the average magnetic pressure. For example, when equipartition is assumed the 
magnetic field energy density clearly dominates the turbulent energy density outside
of the inner kpc of NGC~6946 
\citep{bec07}. This is very puzzling, because the turbulent magnetic field is believed
to be connected to turbulent gas motions. When reducing the magnetic 
energy density by a factor of four, however, it is almost equal to the turbulent energy density.

\section{Rotation measures of galaxies}

We simulated two samples of galaxies at 4.8~GHz ($\lambda$6.25~cm) and 8.4~GHz 
($\lambda3.6$~cm), which are standard frequencies for nearby galaxy 
observations. RM maps of nearby galaxies are mainly calculated from 
polarisation measurements at these frequencies \citep[e.g.][]{bec07}. It is 
generally assumed that the galaxies are Faraday thin at these frequencies, 
which means that the observed polarised emission originates from the same 
volume. We placed the simulated galaxies at a distance of 5~Mpc, corresponding 
to a maximum size of about $23\arcmin\times23\arcmin$, when the galaxies are 
seen face-on. The resolution or the grid size of the simulations is $0\farcm3$ 
or about 440~pc, which is comparable to observations and provides some details 
of the large-scale emission from the galaxies. 

We simulated maps for all Stokes parameters and from them calculated 
polarisation angles ($PA$) at both frequencies. From the polarisation angle 
maps, we obtained RM for each individual pixel as, 
\begin{equation}\label{rm_def}
{\rm RM}=\frac{PA_1-PA_2}{\lambda_1^2-\lambda_2^2},
\end{equation}
where the subscripts stand for the two bands, e.g. $\lambda$ = 6.25~cm and 
$\lambda$ = 3.6~cm. The $n\pi$-ambiguity of the polarisation angles at these 
two bands corresponds to a RM of 1194~rad~m$^{-2}$. We assume that $n$=0 and 
the RM values are smaller than the ambiguity, which is certainly true except 
for the centre regions. 

The Faraday depth (FD) of a galaxy in rad~m$^{-2}$ is defined as,
\begin{equation}
{\rm FD}=0.81\int _{\rm far}^{\rm near} n_e B_\parallel {\rm d}l,
\end{equation}
where $n_e$ is the thermal electron density in cm$^{-3}$ and $B_\parallel$ is 
the magnetic field parallel to the line-of-sight in $\mu$G. The integral is 
calculated along the line-of-sight from the far to the near side of the galaxy. 
It is obvious that FDs are very important to understand the properties of 
magnetic fields and thermal electron densities. In the case that the 
synchrotron-emitting medium and the thermal gas are uniformly mixed, one would 
expect ${\rm FD}=2{\rm RM}$ \citep[e.g.][]{sbs+98}. However, as we will show 
below, the relation does not hold anymore when the frequency-dependent 
polarisation horizon is shorter than the size of a galaxy. 

The RM and FD maps for a nearly face-on ($i=20\degr$) and a nearly edge-on 
($i=80\degr$) galaxy are shown in Fig.~\ref{rmmap}. In general, both RM and 
FD maps show the same intensity pattern, which means that RM maps based on 
observations correctly indicate the orientation of the large-scale field of a 
galaxy. The field reversals in the Milky Way Galaxy and the opposite sign of 
the halo magnetic field above and below the Galactic plane are reflected in the 
RM maps.   

\begin{figure*}[!htbp]
\centering
\resizebox{0.4\textwidth}{!}{\includegraphics[angle=-90]{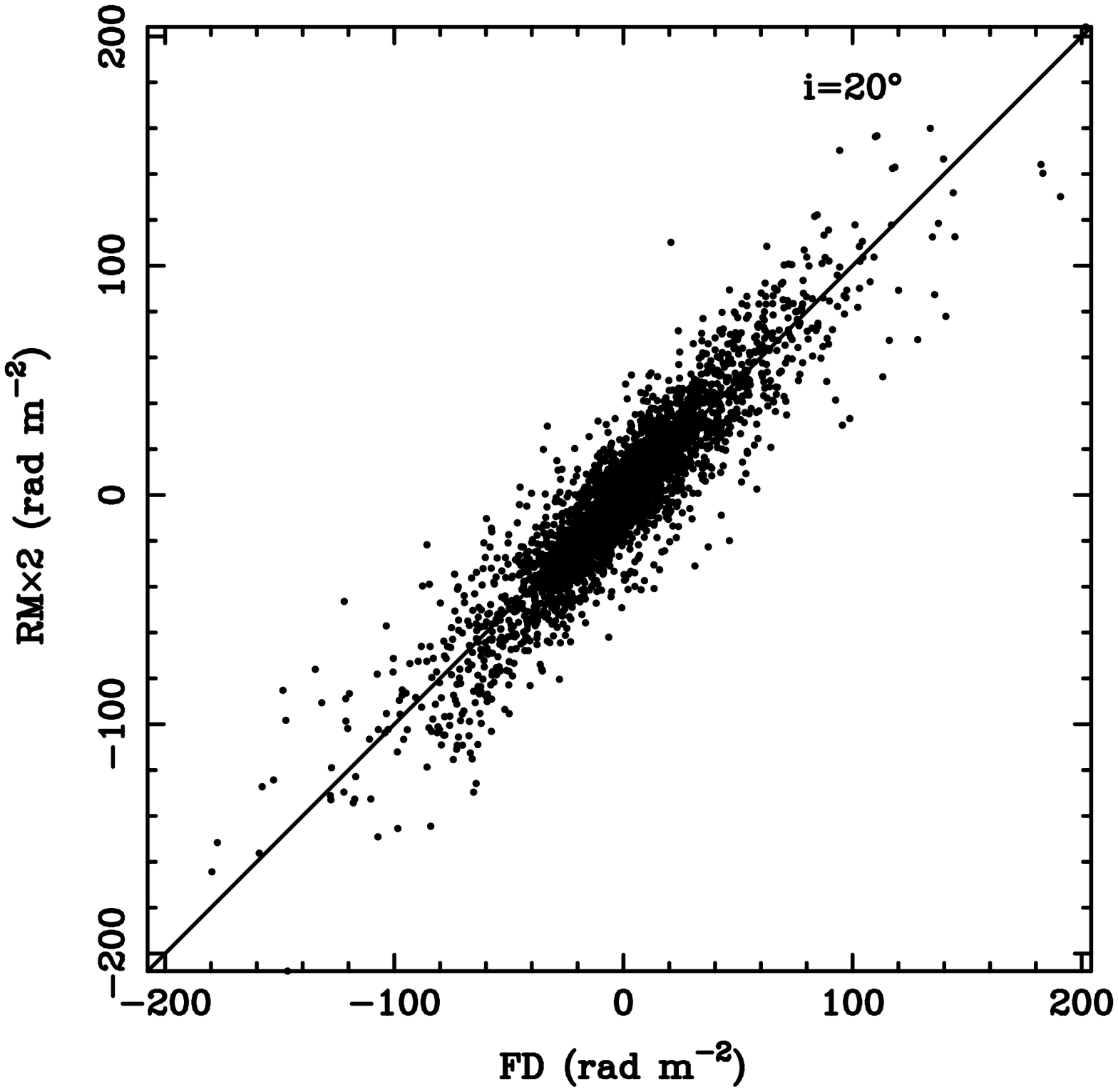}}
\hspace{1cm}
\resizebox{0.4\textwidth}{!}{\includegraphics[angle=-90]{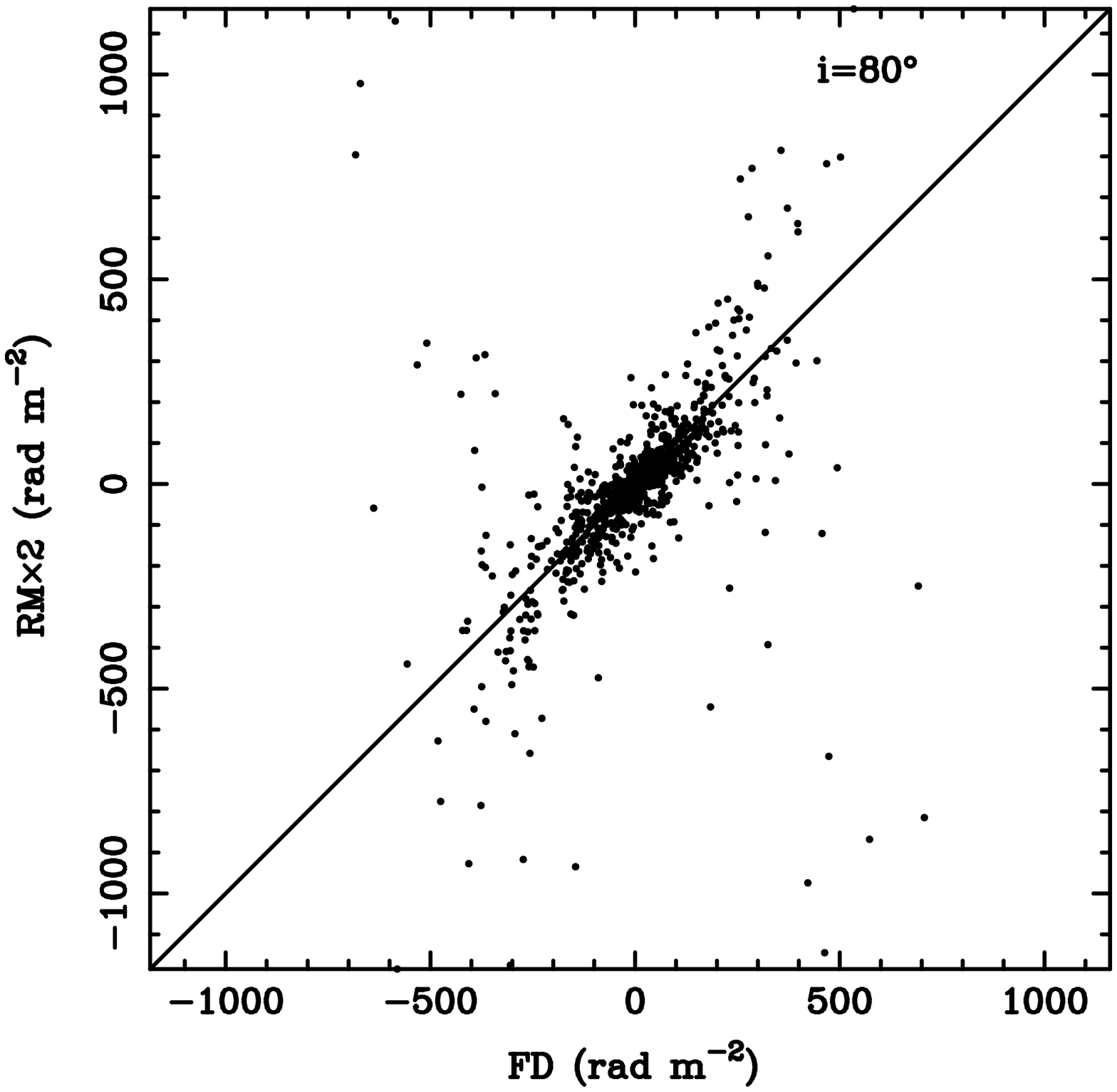}}
\resizebox{0.4\textwidth}{!}{\includegraphics[angle=-90]{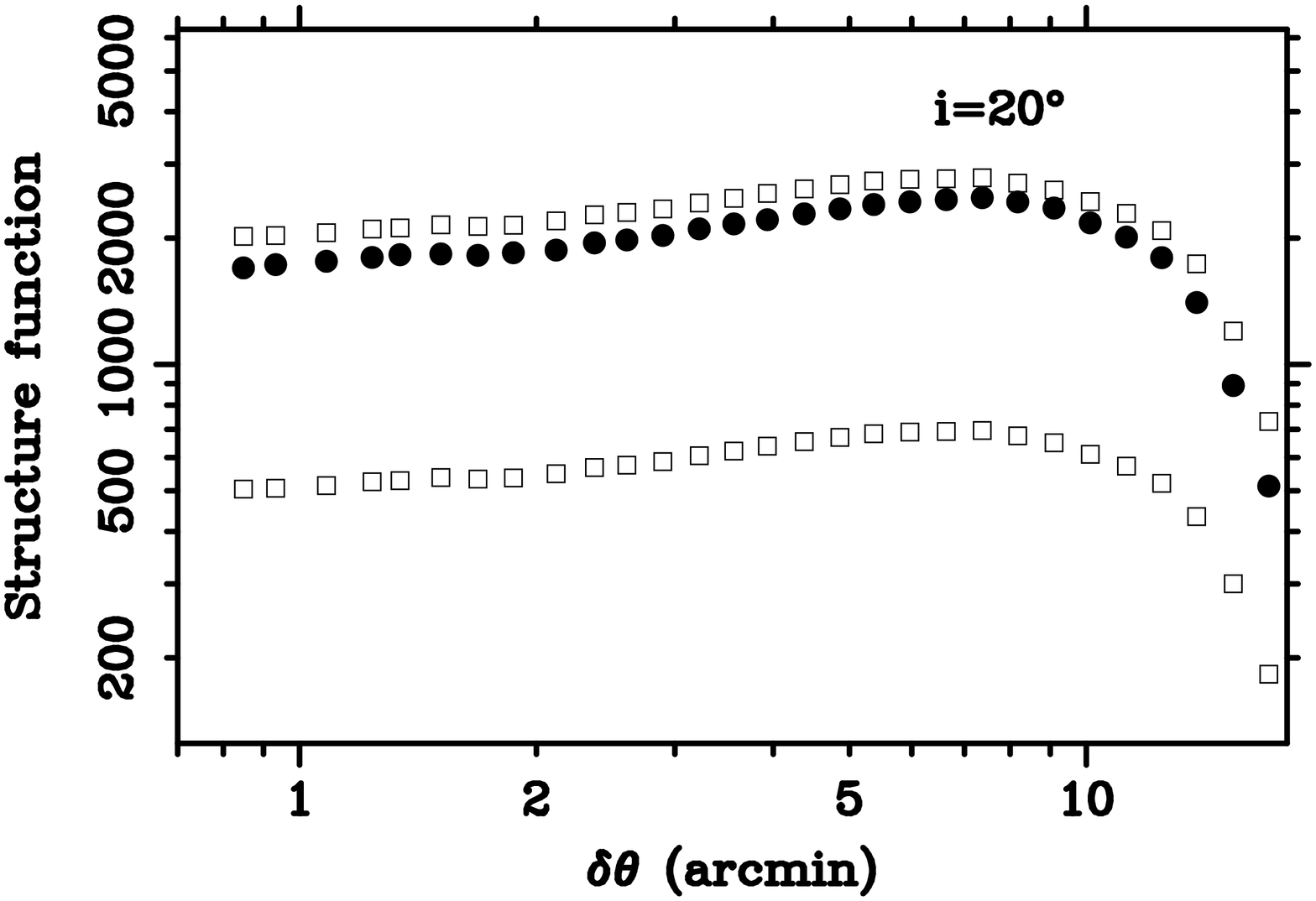}}
\hspace{5mm}
\resizebox{0.4\textwidth}{!}{\includegraphics[angle=-90]{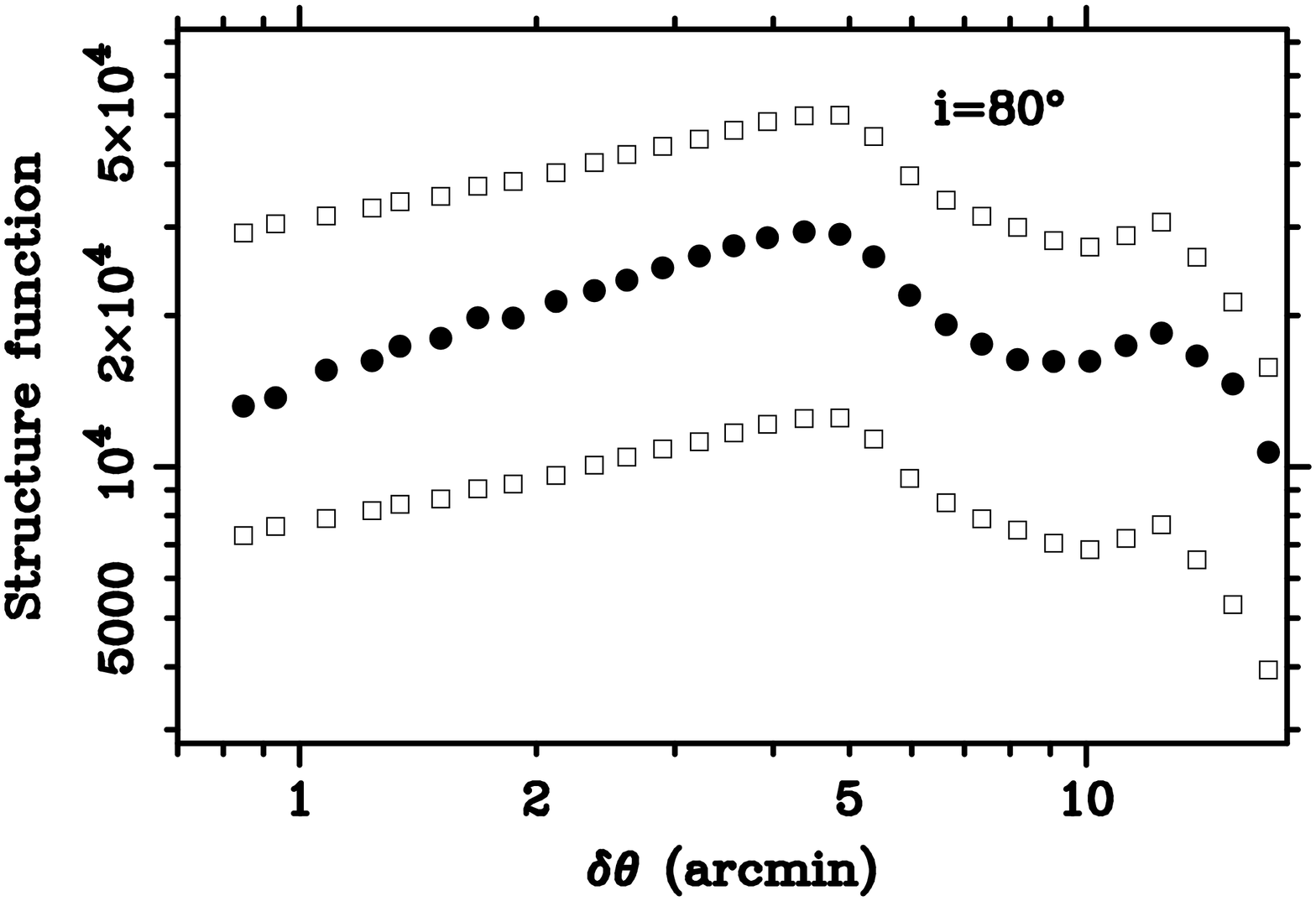}}
\caption{RMs times a factor of two versus FDs ({\it top row}). The solid line 
indicates the expectation for uniformly mixed thermal and synchrotron emission. 
The structure functions ({\it bottom row}) from the FD maps are shown by black 
dots, those from RM maps and RMs scaled by a factor of two by squares.}
\label{fdrm}
\end{figure*}

The RMs multiplied by a factor of two were compared to the corresponding FDs in 
Fig.~\ref{fdrm}. For nearly face-on galaxies ($i=20\degr$) both distributions 
are basically consistent with each other, although a small amount of scatter is 
visible. For nearly edge-on galaxies ($i=80\degr$), however, the differences 
are significant. Large deviations from FD (Fig.~\ref{fdrm}) are seen for large 
FD values originating in the disk. The reason is that the simulated 4.8~GHz and 
8.4~GHz polarised emission does not originate from the same volume. The 
polarisation horizon at 4.8~GHz is smaller than that at 8.4~GHz. 
Higher-frequency observations are needed to overcome this effect and provide 
RMs proportional to FDs.   

Structure functions \citep[e.g.][]{sr09} for RMs and FDs were calculated 
following 
\begin{equation}
\mathcal{D}(\mathbf{\delta\theta})=<[{\rm RM}(\mathbf{\theta})-
{\rm RM}(\mathbf{\theta}+\mathbf{\delta\theta})]^2>,
\end{equation}
where $<\ldots>$ stands for ensemble average and $\mathbf{\delta\theta}$ is the 
angular separation. The results are shown in Fig.~\ref{fdrm}. The structure 
functions for RM, when scaled by a factor of two, is equivalent to shifting 
constantly the structure functions for the original RM upwards along the 
ordinate axis in Fig.~\ref{fdrm}. For galaxies with high inclinations, the 
slopes of the structure functions differ for RMs and FDs, while for 
low-inclination galaxies RMs mimic FDs quite well after being multiplied by a 
factor of two.  

In summary, for face-on galaxies RMs derived from 4.8~GHz and 8.4~GHz 
polarisation measurements can be safely used to infer magnetic fields after 
being scaled by a factor of 2. For edge-on galaxies or highly inclined galaxies,
 however, RMs are only reliable in the outskirts of these galaxies, but still 
indicate the orientation of the magnetic field direction. 

In Fig.~\ref{rmmap} we show maps of the depolarisation between the two 
simulated maps at 4.8~GHz and 8.4~GHz, which show clear differences related to 
their inclination. For the definition of depolarisation (DP$_{\rm CX}$) we 
follow \citet{hbkd11} as, 
\begin{equation}
{\rm DP_{CX}}=100 \frac{PI_{\rm 4.8\,GHz}}{PI_{\rm 8.4\, GHz}}
\left(\frac{8.4}{4.8}\right)^\beta,
\end{equation}
 where $\beta=\alpha-2=-3$ is the spectral index for brightness temperature. 
A value of ${\rm DP_{CX}}=100$ means no depolarisation, and ${\rm DP_{CX}}=0$ means total 
depolarisation. In the depolarisation map for the galaxy with inclination 
$i=20\degr$, the values exceed 92\% everywhere. For the nearly edge-on galaxy 
with $i=80\degr$, the DP$_{\rm CX}$ along the major axis is small and the values close to 
the centre are about 40\%. Although our model does not include discrete thermal 
complexes such as \ion{H}{II} regions, which are certainly present along the 
spiral arms, the depolarisation maps show a patchy morphology. This results 
from the various random magnetic field components along line-of-sights. The 
highly inclined galaxy NGC~253 with $i=78\fdg5$ was recently studied by 
\citet{hbkd11} using polarisation observations at 4.8~GHz and 8.4~GHz. Their 
depolarisation map shows similar values as our results, although their 
distribution of depolarisation patches is more complex than ours. The main 
reason is the missing discrete source complexes in our model.

We also calculated the depolarisation (DP$_{\rm LC}$) between 1.4~GHz and 
4.8~GHz (Fig.~\ref{rmmap}). For low inclination galaxies, DP$_{\rm LC}$ is 
around 40\% within $5\arcmin$ and there is a slight asymmetry along the minor axis 
direction, which is caused by the toroidal halo field. We note that some 
galaxies, for instance NGC~6946, show a clear asymmetric polarisation at 
frequencies around 1.4~GHz with complete depolarisation towards a large patch 
\citep{bec07,hbe09}. This might indicate a more complex halo field configuration 
than in the Milky Way. A quadrupole field configuration was discussed by 
\citet{bhb10}, but asymmetries could also be ascribed to some large-scale 
distortions from a regular configuration. In our Galaxy, discrete large RM or
polarisation features exist, for instance the well pronounced large Fan region 
\citep[e.g.][]{wlrw06}, which are commonly considered to be local. If such regions
reside out of the Galactic plane or in the halo, they would make the polarisation
properties of our Galaxy asymmetric and thus similar to others. More investigations 
are needed to improve Galactic emission models.

\section{Integrated polarisation properties of Milky Way type galaxies}

Spiral galaxies are believed to occupy a large fraction of the polarised radio 
sources at 1.4~GHz at $\mu$Jy flux densities \citep{stil07}. However, they are 
difficult to be resolved individually even with instruments such as ASKAP 
\citep{jbb+07} or the future SKA. It is therefore important to investigate the 
polarisation properties of unresolved spiral galaxies. 

\begin{figure*}[!htbp]
\centering
\resizebox{0.9\textwidth}{!}{\includegraphics[angle=-90]{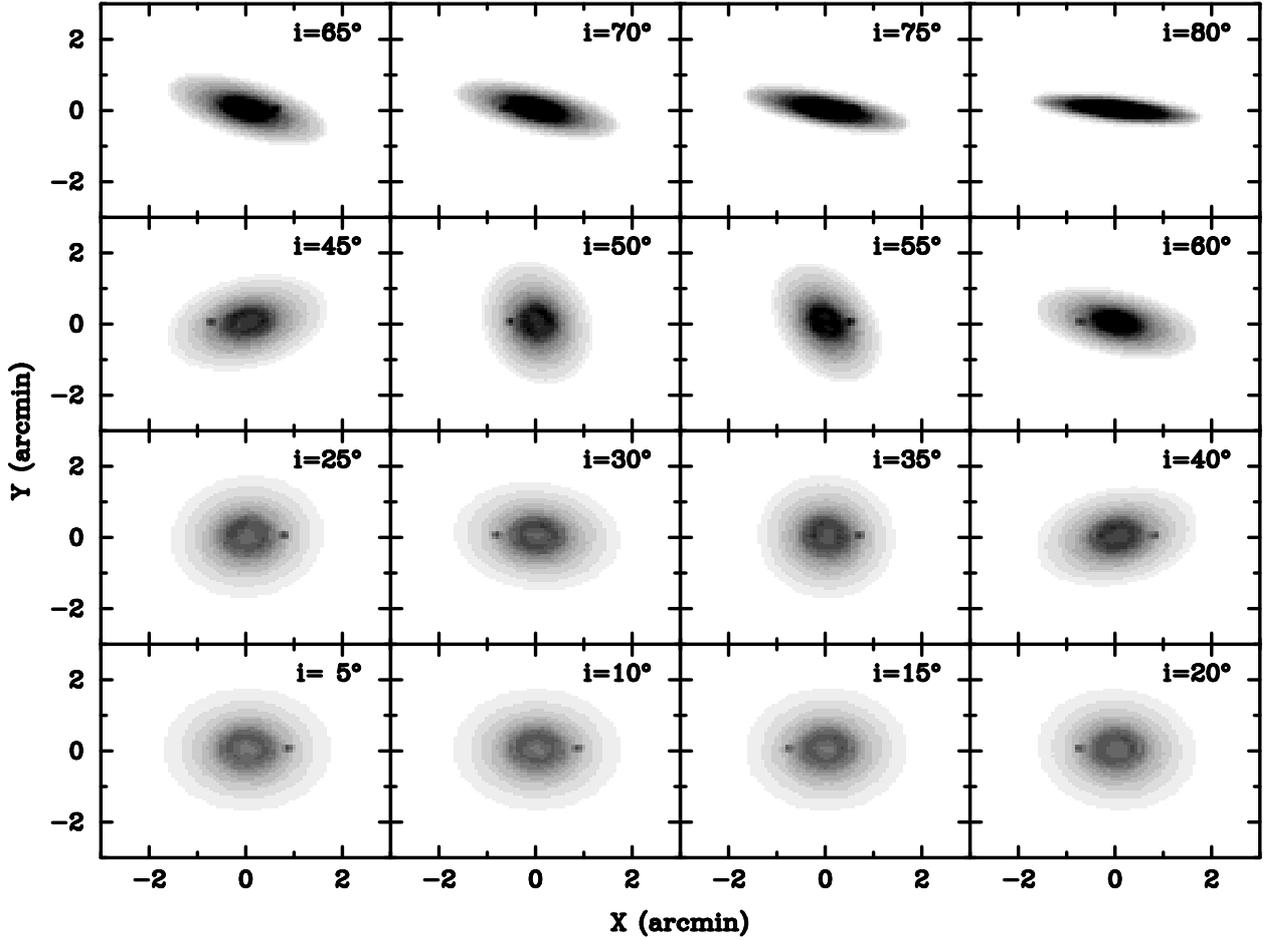}}
\caption{Sample of Milky-Way-like galaxies at 1.4~GHz in total intensity. 
The inclination 
angles are marked. The position of the Solar System (local enhancement) is 
indicated by a black dot. The angular resolution is $0\farcm3$. The gray scale 
runs from 0 to 1.5~K.}
\label{egsL}
\end{figure*}

We simulated 1000 Milky-Way-like galaxies by randomly 
setting $\theta_1$ and $\theta_2$ 
(see Sect.~2.2) at each of the five frequencies: 1.4~GHz, 2.7~GHz, 4.8~GHz, 
8.4~GHz, and 22~GHz. The distance was set to 10~Mpc and the angular resolution to 
$18\arcsec$ or 0.87~kpc. We have tried simulations at higher angular 
resolutions and the results are similar.
A sample of Milky-Way-like galaxies at 1.4~GHz is shown in 
Fig.~\ref{egsL}.   

\begin{figure}[!htbp]
\centering
\resizebox{0.48\textwidth}{!}{\includegraphics[angle=-90]{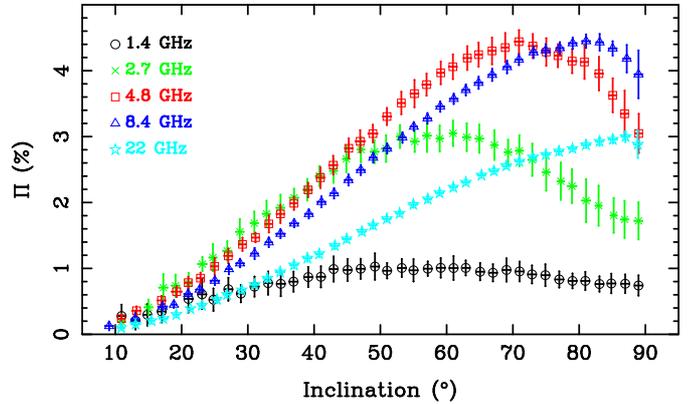}}
\caption{Polarisation percentages of unresolved Milky Way galaxies versus 
inclination angles at several frequencies as indicated.}
\label{pci}
\end{figure}

For each object we integrated the 
$I$, $U$ and $Q$ maps and obtained the polarisation intensity 
$PI=\sqrt{U^2+Q^2}$ and the polarisation percentage $\Pi=PI/I$. We binned the 
data according to their inclination angles with a step interval of $2\degr$. 
The average percentage polarisation as a function of inclination is shown in 
Fig.~\ref{pci}.

Face-on galaxies show a small degree of polarisation because of the assumed 
symmetry of the magnetic field in the disk, which strongly cancels the $U$ and 
$Q$ components when integrated. This is common for all frequencies 
(Fig.~\ref{pci}). Faraday effects become highly significant for edge-on 
galaxies because of the large path-length along the disk emission, which 
renders the polarisation percentage low. Therefore the maximum polarisation 
percentage is expected in between, around $i=60\degr$ at 2.7~GHz, around 
$i=70\degr$ at 4.8~GHz and around $i=80\degr$ at 8.4~GHz. The shift stems from 
the reduction of Faraday depolarisation towards higher frequencies. At low 
frequencies such as 1.4~GHz, however, Faraday effects are strong irrespective 
of inclinations, which always yields small polarisation percentages. At high 
frequencies, the thermal fraction in a galaxy gets larger, which consequently 
reduces the polarisation percentage despite reduced Faraday depolarisation and 
explains the lower percentage polarisation at 22~GHz when compared with those 
at 8.4~GHz.

The polarisation percentage at 4.8~GHz (Fig.~\ref{pci}) peaks at the same 
inclination range as that found by \citet{skbt09} or \citet{sti09}, who used 
a geometrical 
galaxy model. The models by \citet{skbt09} all predict zero polarisation 
percentage for edge-on galaxies, which is not found in our simulations. The 
internal depolarisation for edge-on galaxies was probably overestimated by 
\citet{skbt09}. In those cases, the polarisation horizon is small 
\citep[e.g.][]{srh+11} and much shorter than the diameter of the disk. 
However, the disk size was used by \citet{skbt09} to derive the RM 
fluctuations, which certainly results in too large fluctuation and hence 
depolarisation. In fact, most of the edge-on galaxies observed at 4.8~GHz 
show considerable polarisation percentages (see Fig.~1 in \citealt{skbt09}), 
and agree with our simulations (Fig.~\ref{pci}).

\begin{figure}[!htbp]
\centering
\resizebox{0.45\textwidth}{!}{\includegraphics[angle=-90]{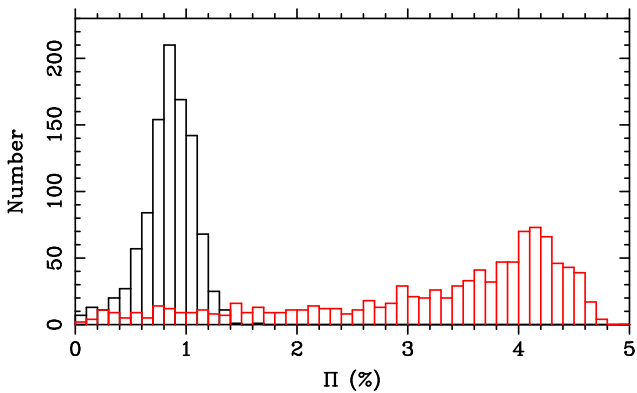}}
\caption{Histogram of polarisation percentages at 1.4~GHz (black) and 4.8~GHz 
(red) for 1000 simulated Milky Way galaxies.}
\label{pchist}
\end{figure}

The distribution of polarisation percentages for 1000 simulated galaxies
at 1.4~GHz and 4.8~GHz are presented in Fig.~\ref{pchist}, which peaks at about 
0.8\% at 1.4~GHz and 4.2\% at 4.8~GHz. 
In our distributions, the relative number of galaxies at zero 
polarisation percentage is very small, which is clearly opposite to that in 
the histograms by \citet[][their Fig.~6]{skbt09}.  
This difference is not caused by varying the parameters in the model of 
\citet{skbt09}, and should be explained in the following way: To 
ensure a spacial uniform distribution of galaxies, the probability to observe a 
galaxy at inclination between $i$ and $i+{\rm d}i$ is proportional to $\sin i$ 
\citep[e.g.][]{skbt09}, meaning the relative number of edge-on galaxies is 
maximal. Provided that edge-on galaxies have zero percentage as modelled by 
\citet{skbt09}, it follows that there is still a significant number 
of galaxies at zero percentage in their distributions.

We note that the polarisation percentages can be as high as about 20\% in the 
models by \citet{skbt09} with varying magnetic field properties. In 
this paper, we simulate Milky-Way-like galaxies with the same 
properties of the interstellar medium, as they were derived from 
observations \citep{srwe08}, and simply rotate the Milky Way Galaxy to 
realise different inclinations. This sample should therefore be considered
as a subset of spiral galaxies.  
In future when more data are available from 
e.g. GALFACTS \citep{ts10}, we will try to represent general galaxies by 
modifying the Milky Way model accordingly. 

\begin{figure}[!htbp]
\centering
\resizebox{0.45\textwidth}{!}{\includegraphics[angle=-90]{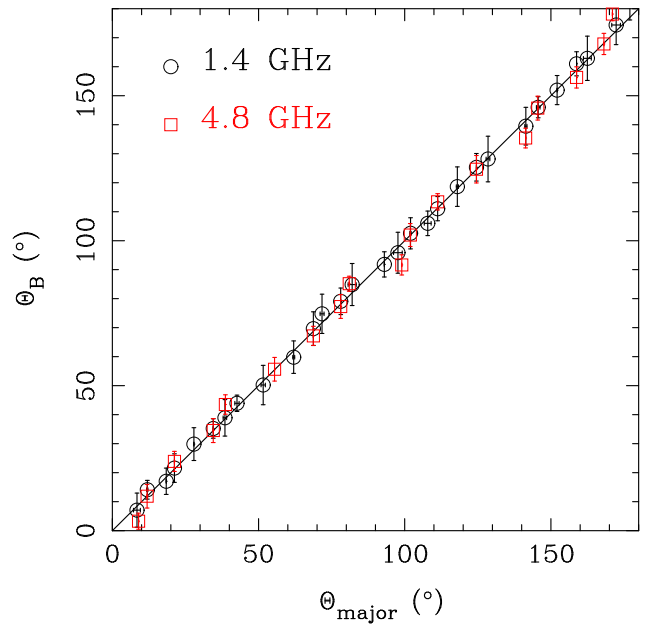}}
\caption{B-vectors ($PA+90\degr$) versus position angles of the major axes from 
simulated Milky Way galaxies at 1.4~GHz and 4.8~GHz.}
\label{pak}
\end{figure}

The observed polarisation from  unresolved galaxies results from averaging the 
projection of their azimuthal magnetic field components on the plane of the 
sky. Therefore we expect an alignment between the B-vectors ($\Theta_B$) and 
the position angles of the major axes ($\Theta_{\rm major} $), which is indeed 
seen from the present simulations as shown in Fig.~\ref{pak}. This means that 
polarisation angles of galaxies, when calculated from their integrated emission,
 do not vary with frequencies, implying that these galaxies do not show 
intrinsic RMs. This result has already been reported by \citet{skbt09} from 
their simulations and is now confirmed. This characteristic makes unresolved 
spiral galaxies with optically measured position angles ideal background 
polarised sources to study Galactic and intergalactic large-scale magnetic 
fields with instruments like the future SKA.

\section{Summary}
The Milky Way Galaxy is widely considered as a typical spiral galaxy. We used 
the observationally well constrained Galactic 3D-emission model 
\citep{srwe08,sr10} to simulate emission of galaxies by putting the Galaxy 
away at any distances and rotating it by any angle.

In our models, the scale-height of the synchrotron emission from the Milky Way 
is much smaller than that assumed in the early models \citep{pko+81,bkb85}. At 
that time, the evidence for the local synchrotron excess was not well 
determined and therefore not taken into account. We fit to our Milky Way model, 
when seen edge-on, an exponential scale-height of 0.74~kpc. 

We tested a current analysis method to derive the scale-height of highly
inclined galaxies. We find that for Milky Way galaxies the scale-height 
apparently gets larger with decreasing inclination. For inclinations larger 
than $80\degr$, however, the increase is below 10\%. The total magnetic field 
estimate based on the equipartition assumption is about twice that of the Milky
Way Galaxy model.
   
Polarisation observations of highly inclined galaxies when used to derive the 
distribution of RMs are not reliable for their inner parts. At frequencies of 
4.8~GHz or lower, they are not Faraday thin anymore. The polarised emission 
from two frequency observations used to calculate RMs originates from different 
depths. This is also reflected in the slope of the structure functions, which 
clearly differ for RMs and FDs for galaxies nearly seen edge-on. 

We simulated a large sample of distant unresolved galaxies as it will be 
accessible to the future SKA and its path-finders, and analysed the 
polarisation properties of this sample for a large frequency range. This 
analysis follows that by \citet{skbt09} of a large spiral galaxy sample based 
on their model of polarised emission. We qualitatively confirm the 
\citet{skbt09} results, 
where the percentage polarisation drops to zero for face-on galaxies at all 
frequencies. For edge-on galaxies \citet{skbt09} found almost no percentage 
polarisation, while we note a moderate polarisation percentage decrease. 
The peak percentage polarisation of the galaxy sample peaks at about 4.2\% at 
4.8~GHz. At 1.4~GHz we find an average percentage polarisations of about 0.8\%. 
We confirm the previous finding by \citet{skbt09} that unresolved spiral 
galaxies have a very well aligned polarisation angle in the direction of the 
galaxies major axis, which makes them ideal sources to study foreground Faraday 
rotation effects on Mpc-scales. 

\acknowledgements{X.S. was supported by the Australian Research Council 
through grant FL100100114. X.S. thanks for financial support by the MPG and by
Prof. M. Kramer during his stay at MPIfR, where most of this work was done.
We thank Patricia Reich and Bryan Gaensler for thorough reading of 
the manuscript and also the anonymous referee for helpful comments.}

\bibliographystyle{aa}
\bibliography{/import/fenway2/xhsun/bibtex.bib}

\end{document}